\documentclass[10pt]{article}
\usepackage[OE]{express}
\usepackage{multirow,amssymb,amsbsy,amsmath,epstopdf}

\begin{document}
\title{Experimental nonlocal steering of Bohmian trajectories}

\author{Ya Xiao,\authormark{1,2} Yaron Kedem,\authormark{3} Jin-Shi Xu,\authormark{1,2,4} Chuan-Feng Li,\authormark{1,2,5} and Guang-Can Guo\authormark{1,2}}

\address{\authormark{1}CAS Key Laboratory of Quantum Information, University of Science and Technology of China, Hefei 230026, People's Republic of China\\
\authormark{2}Synergetic Innovation Center of Quantum Information and Quantum Physics, University of Science and Technology of China, Hefei 230026, People's Republic of China\\
\authormark{3}Nordita, Center for Quantum Materials, KTH Royal Institute of Technology and Stockholm University, Roslagstullsbacken 23, 10691 Stockholm, Sweden\\
\authormark{4}jsxu@ustc.edu.cn\\
\authormark{5}cfli@ustc.edu.cn
}



\begin{abstract}
Interpretations of quantum mechanics (QM), or proposals for underlying theories, that attempt to present a definite realist picture, such as Bohmian mechanics, require strong non-local effects. Naively, these effects would violate causality and contradict special relativity. However if the theory agrees with QM the violation cannot be observed directly. Here, we demonstrate experimentally such an effect: we steer the velocity and trajectory of a Bohmian particle using a remote measurement. We use a pair of photons and entangle the spatial transverse position of one with the polarization of the other. The first photon is sent to a double-slit-like apparatus, where its trajectory is measured using the technique of Weak Measurements. The other photon is projected to a linear polarization state. The choice of polarization state, and the result, steer the first photon in the most intuitive sense of the word. The effect is indeed shown to be dramatic, while being easy to visualize. We discuss its strength and what are the conditions for it to occur.
\end{abstract}



\section{Introduction}

The non-local features of quantum mechanics have intrigued physicists for over a century. One of the first attempts to formalize non locality as a physical effect was the de Broglie-Bohm theory, or Bohmian Mechanics (BM) \cite{Bhom1,Bhom2}. There, the wavefunction is treated as a potential, affecting the trajectories of particles. The particles themselves are localized as realistic objects, in the sense that they show no interference effect. The wavefunction, on the other hand, is a non-local object so any change to it, even if it is due to a well localized event, is global. The implication is that performing some action in one position can immediately modify the trajectory of a particle in another, causally disconnected, position. In recent years, it has been realized that, with the help of weak measurement \cite{Aharonov1988,Kofman2012,Dressel2014,Tamir2013,Aharonov2014}, it is possible to observe the average trajectory of a Bohmian particle \cite{Kocsis2011}. Even more recently, a proposal to observe non local effects within BM was put forward \cite{Braverman2013} and later realized \cite{Mahler2016}.

Another manifestation of quantum non locality is the concept of quantum steering, which was first proposed by Schr\"{o}dinger \cite{S1935,S1936} and was recently defined rigorously \cite{Wiseman2007}. Obtaining a certain result in a measurement performed in one location, which depends on the local choice of the measurement basis, can determine the measurement statistics in another location. Thus the state of the second location is ``steered'' by the results obtained in the first location. The idea was investigated widely both theoretically \cite{Bowles2014,Quintino2014,Piani2015,
Kogias2015} and experimentally \cite{Saunders2010,Sun2014,Sun2016,Wollmann2016,Xiao2017}.

Combining these two concepts we get a physical picture that is non local in a deep sense: the measurement done in one place, immediately deflects the trajectories of a particle that can be far away, in a causally disconnected region. Indeed, the final results agrees with the prediction of quantum mechanics, and no superluminal information transfer is possible. However, the underlying picture, which is as valid as any other formulation of quantum mechanics, highlights the inevitable non-local nature of any theory compatible with quantum mechanics.

\section{The steering process in the framework of Bohmian mechanics.}

In this section, we give a full Bohmian mechanics (BM) description of the process of steering. The measurement process is described by a unitary evolution, which splits the wavepacket. We show how this step modifies the velocity of another particle.

The velocity of a Bohmian particle $j$ is given by
\begin{equation}\label{velocity}
v_j = \text{Re} \dfrac{\psi^{*}P_j \psi}{m_j  \vert\psi\vert^{2}},
\end{equation}
where $P_j =  -i \dfrac{\partial}{\partial q_j} $  is the quantum mechanical momentum operator for the specific particle, $m_j $ is its mass and $\psi = \psi(q_1,q_2,...) $ is the global wave function evaluated at the (Bohmian) positions of all particles $q_i$. We consider a system of two particles and two additional degrees of freedom. The additional degree of freedom can be any discrete level system, such as spin or polarization. Below we refer to it as spin, for brevity. The initial wavefunction is
\begin{equation}\label{psi0}
\psi = \dfrac{1}{\sqrt{2}}f(q_B) g(q_A) \left(\vert{\uparrow}\rangle_B \vert{\uparrow}\rangle_A + \vert{\downarrow}\rangle_B \vert{\downarrow}\rangle_A \right),
\end{equation}
where $q_{A(B)}$ is the position of particle $A (B)$, $f$ and $g$ are localized wavepackets, at different locations.  $\vert{\uparrow}\rangle_{A(B)}$ and $\vert{\downarrow}\rangle_{A(B)}$ describe the spin. In this example, an operation (measurement) done in the location of particle $A$ will modify the velocity of particle $B$. As seen from Eq. (\ref{psi0}), initially, the entanglement is only in the spins and not in the positions of the particles. So, we should transfer the entanglement by splitting the wavepacket of particle B according to its spin.
This is done by the unitary operator
$
U_1 = e^{- i d P_B \vert{\uparrow} \rangle_B \langle{\uparrow}\vert_B },
$
where $ P_B$ is the momentum operator for particle $B$ and $d$ is the distance of the splitting. After this operation, the wavefunction is
\begin{equation} \label{psi1}
\psi = \dfrac{1}{\sqrt{2}}g(q_A) \left(f(q_B+d) \vert{\uparrow}\rangle_B \vert{\uparrow}\rangle_A + f(q_B ) \vert{\downarrow}\rangle_B \vert{\downarrow}\rangle_A \right).
\end{equation}
At this point, the velocity of particle $B$, given by inserting Eq. (\ref{psi1}) into Eq.  (\ref{velocity}), does not depend on the position $ q_A$. One can view it as a trace over the spin, or as a conditional average of the two wave packets \cite{gilli}. Note that any phase associated with the different component in (\ref{psi1}) would not affect the velocity, since the entanglement prevents them from interfering.

The measurement on particle A, is done by applying a similar operation,
$
U_2 = e^{- i L P_A \vert{\theta}\rangle_A \langle{\theta}\vert_A } $
where $ P_A$ is the momentum operator for particle A, $L$ is a distance much bigger than the width of $g(q_A)$ and the state
\begin{align} \label{theta}
\vert{\theta}\rangle_A = \cos \theta \vert{\uparrow}\rangle_A - \sin \theta \vert{\downarrow}\rangle_A,
\end{align}
is chosen by the experimentalist at the location of particle $A$. The wavefunction after this operation is
\begin{align} \label{psi2}
\psi = {g(q_A +L ) \over \sqrt{2}} &\left[  f(q_B+d) \cos(\theta) \vert{\uparrow}\rangle_B -  f(q_B ) \sin (\theta) \vert{\downarrow}\rangle_B \right] \vert{\theta}\rangle_A  \nonumber \\
+ {g(q_A )\over \sqrt{2}} &\left[ f(q_B+d)  \sin (\theta)\vert{\uparrow}\rangle_B +  f(q_B ) \cos (\theta)\vert{\downarrow}\rangle_B \right] \vert{\bar{\theta}}\rangle_A,
\end{align}
where $\vert\bar{\theta}\rangle_A = \sin \theta \vert{\uparrow}\rangle_A + \cos \theta \vert{\downarrow}\rangle_A$ is the state orthogonal to $\vert{\theta}\rangle_A$.
Even though Eq. (\ref{psi2}) is written as a superposition, in the quantum mechanical sense, in the context of BM it is not. When calculating the velocity of particle $B$ by inserting (\ref{psi2}) into (\ref{velocity}), $\psi$ is evaluated at the Bohmian positions of all particles. Since, for any $q_A$, either $g(q_A +L )$ or $g(q_A )$ vanish (for large enough $L$) only one of the lines in Eq. (\ref{psi2}) is relevant. Thus the operation of $U_2$, which can be done promptly, changes the velocity of particle $B$, i.e. it is being steered.

\section{Experimental setup and results}

\begin{figure}[!htb]
\begin{center}
\includegraphics[width=1\columnwidth]{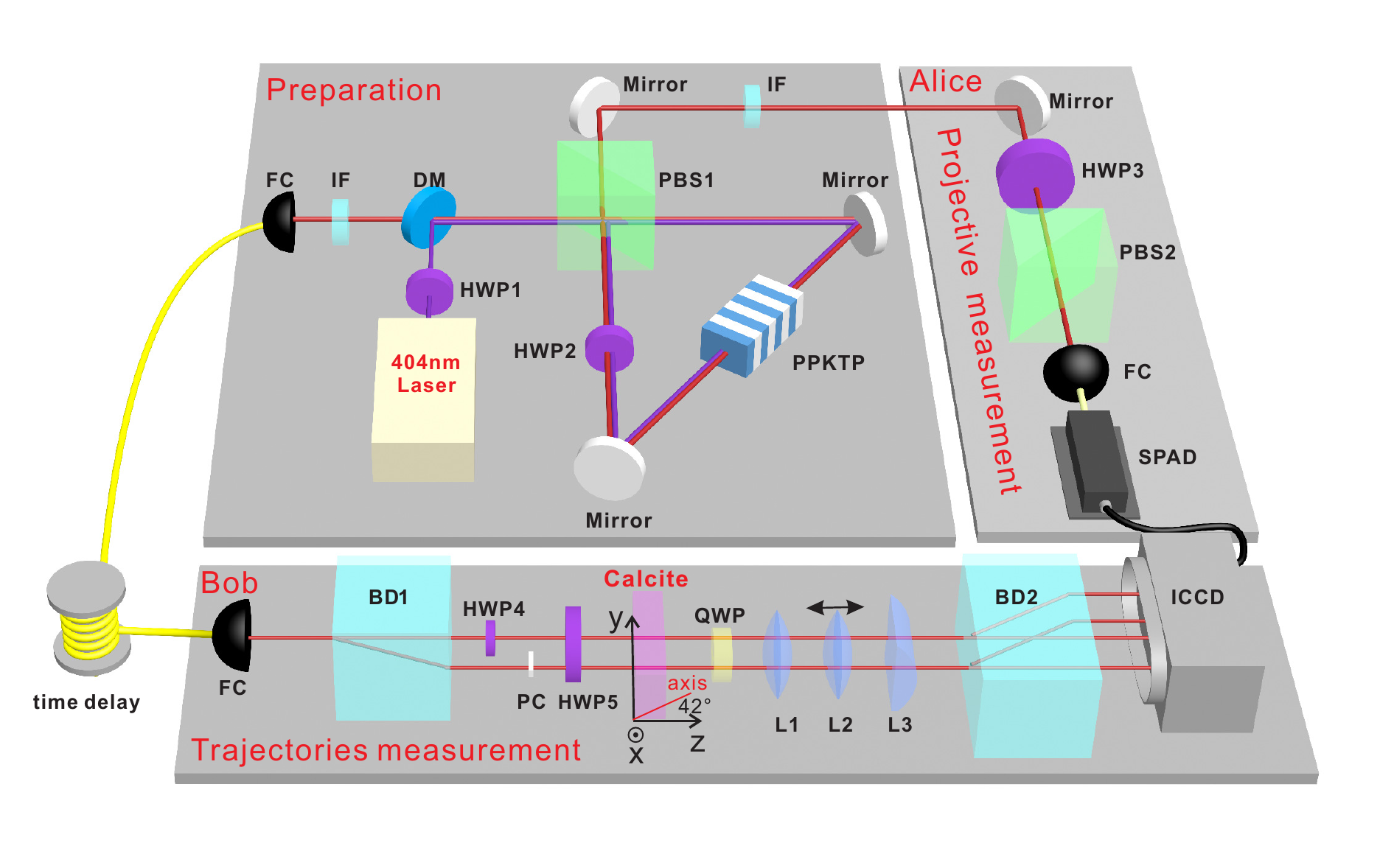}
\caption{Experimental setup. Preparation of polarization-entangled states: An ultraviolet laser is reflected by a dichroic mirror (DM) to pump the PPKTP crystal located in a Sagnac interferometer. Dual-wavelength half-wave plates (HWP1 and HWP2) are used to rotate the polarizations of the pump laser and the down-converted photons which are separated by the dual-wavelength polarizing beam splitter (PBS1). Interference filters (IF) are used in both paths. Projective measurement: The polarization of photon A (Alice's side) is projected using HWP3 and PBS2. It is then coupled by a fiber coupler (FC) and detected by the single-photon avalanche detector (SPAD) with the electric signals using as a trigger. Trajectories measurement:  Photon B (Bob's side) is coupled to an 85 m long single-mode fiber and then separated into two paths by a beam displacer (BD1). HWP4 is used to make the polarization of the two beams identical and birefrigent crystals (PC) are used to compensate the difference in the optical length. The polarization of the two beams is rotated by HWP5 and the photon goes through a thin calcite crystal to perform weak measurement. A quarter wave plate (QWP) and BD2 are used to detect the polarization of the photon. Lens L1 (plano-convex), L2 (aspherical, movelable) and L3 (plano-convex cylindrical) are used to image different planes in the ICCD camera which is triggered by the SPAD.}
\label{setup}
\end{center}
\end{figure}

We performed an experiment to demonstrate the steering of Bohmian velocities using entangled photons. Figure \ref{setup} shows our experimental setup. A 404 nm laser (Toptica Bluemode) is used to bidirectionally pump a 20 mm long PPKTP crystal located in a polarization Sagnac interferometer \cite{Fedrizzi2007}. Maximally polarization-entangled photon pairs (photon $A$ for Alice's side and $B$ for Bob's side) are generated with the state in the form of  $\vert \psi\rangle= \dfrac{1}{\sqrt{2}}(\vert HH\rangle+\vert VV \rangle) $, where$ |H\rangle $  and $ \vert V\rangle $  are the horizontal and vertical polarizations, respectively. The quality of the entangled photon source is characterized by the entanglement measurement, concurrence \cite{Wootters1998}, which is measured to be 0.964. Photon $B$ is sent to a calcite beam displacer, creating a displacement of 3 mm between the horizontal and vertical components. The resulting wavefunction of the two photons is given by Eq. (\ref{psi1}), where the spin state denotes the polarization $\vert {\uparrow (\downarrow) } \rangle= \vert{H (V)}\rangle$ and $q_{A (B)}$ is the transverse position of photon $A(B)$, which we denote as $x$ for the steered photon $B$. The polarizations of the two parts of photon $B$ are further set to be the same and the difference in their optical path is compensated, such that they can only be distinguishable by the transverse location. Photon $A$ goes through a polarized beam splitter, at an angle $\theta$ with respect to the horizontal axis, such that it is projected to the state $ \vert{\theta}\rangle$ in Eq. (\ref{theta}). The detected signal is used to trigger the detection of photon $B$, which is delayed by an 85-m-long single-mode fiber, to ensure the triggering can be done in time.


The Bohmian velocity can be obtained as
$v(x) = {1 \over m} \text{Re} \left[ P_w(x) \right]$  \cite{Wiseman2007_2,Durr2009}.
$
P_{w}(x) =\dfrac{\langle x\vert P_{w} \vert\psi\rangle}{\langle x \vert\psi\rangle},
$ is the weak value \cite{Aharonov1988} of the momentum operator, where  $ \vert{x}\rangle$ is the state of the particle being at $x$. In the paraxial approximation, the momentum is replaced by the relative transverse momentum $k_x$ and the time is replaced by the longitudinal distance $z$. The velocity is then given by
\begin{equation} \label{vx}
v_x(x,z) = c  {\langle k_{x}\rangle_{w} \over  k},
\end{equation}
where $c$ is the speed of light and $k$ is the magnitude of the momentum.
The measurement of $\langle k_{x}\rangle_{w}$ is performed by placing a piece of 0.7-mm-thick calcite, with its optic axis in the {\it x-z} plane, oriented at  $42^{\circ}$ with respect to the $z$ axis.  This induces a small rotation in the polarization state, which depends on the transverse momentum $\vert{ H} \rangle+ \vert{V}\rangle \rightarrow \vert{ H} \rangle+ e^{i \zeta k_x/k} \vert{V}\rangle $, where $  \zeta =336 $ is a dimensionless coupling strength. This rotation is then measured by a quarter wave plate and a beam displacer (BD2), which deflects each polarization component to its own section in an ICCD camera (Andor iStar 334), where the traverse position $x$ is detected. For each pair of pixels, at position $x$ in the two sections, the weak value of the transverse momentum is obtained by
$\langle k_{x}\rangle_{w}=\dfrac{k}{\zeta} \arcsin(\dfrac{I_{R}-I_{L}}{I_{R}+I_{L}})$,
where $I_{R (L)}$ is count corresponding to right (left) circular polarization. We provide the detailed method to measure the coupling strength and calculate the momentum in the Appendices A and B. A system with three lenses, L1 (plano convex), L2 (aspherical) and L3 (plano-convex cylindrical), allows us to image, at the ICCD camera, planes of varying optical effective distance $z$, by moving L2. 45 imaging planes were measured in the range z=1.492 m to z=4.500 m.

\begin{figure}[!htb]
\begin{center}
\includegraphics[width=1\columnwidth]{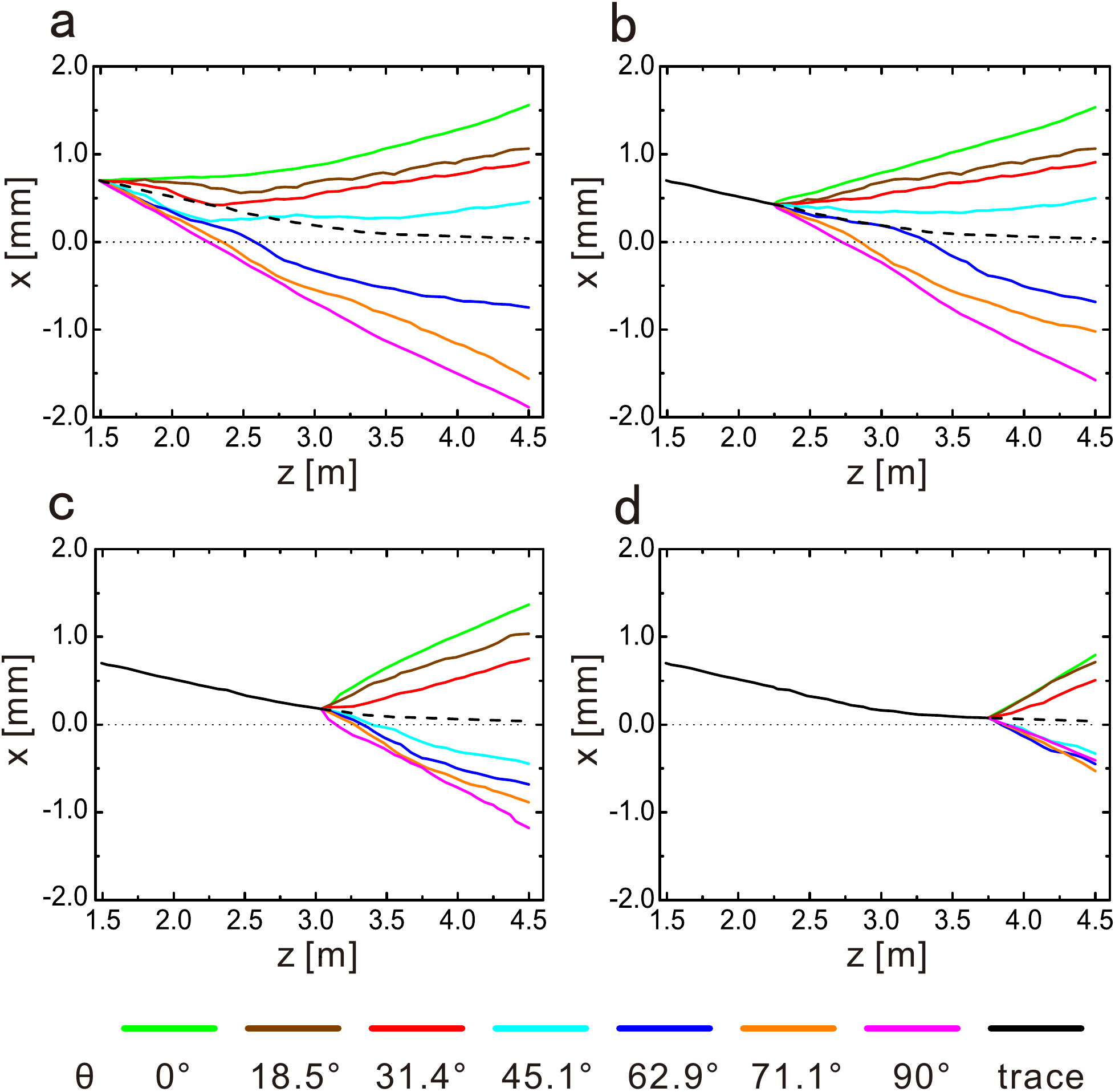}
\caption{Steered Bohmian trajectories. The lines are reconstructed trajectories of photon B, using the measured Bohmian velocities. The trajectories start at one point and follow the non-projected velocity until at some point, a measurement on photon A takes place, and this modifies the velocity of photon B. The state in which photon A is found, determines the new direction for photon B, and from that point we plot the trajectories according to a number of states $ \cos\theta \vert H \rangle - \sin\theta \vert V \rangle $, shown in different colors for different $ \theta $ (the non-projected trajectory is represented by the black dashed line after the point). The different panels represent different times for the measurement on photon A, which correspond to the time photon B has passed a longitudinal distance of (a) $z$=1.492 m, (b) $z$=2.245 m,  (c) $z$=3.038 mm, and (d ) $z$=3.749 m. The black dot line
indicates the center between the two slits.}
\label{steer}
\end{center}
\end{figure}

To reconstruct the average trajectories, we take an initial transverse position, $x_{0}$, at the first imaging plane, $z_0$=1.492 m, and then find the next position at the next plane according to $ x_{j+1}=x_{j}+(z_{j+1}-z_j) v_x(x_j,z_j) /\sqrt{c^{2}-v^{2}_x(x_j,z_j)} $,
where $x_j$ and $z_j$ are the transverse and longitudinal position, respectively. A trajectory, for any choice  of $x_{0}$ is obtained by connecting all the position $x_{j}$ following that choice. According to BM, the particle has a definite trajectory, before and after the projection of photon $A$. At the moment photon $A$ is measured, the velocity of photon $B$ is changed and the trajectory continues in a different direction, according to the result of the measurement. We demonstrate this effect by plotting a trajectory, according to Eq. (\ref{psi1}), up to some point, which represents the time the measurement on photon $A$ took place. The rest of the trajectory is plotted, using the results of a projection, according to Eq. (\ref{psi2}). In Fig. \ref{steer}, we show this for a few points of projection, i.e. a few times of the measurement, and for a number of projection states. The result is a vivid demonstration of how Bohmian trajectories are steered by a measurement done in a remote location. Until the measurement takes place the particle follows a single trajectory. At the time the measurement occurs, its pilot-wave, or ``quantum potential'' is altered and so is its velocity. The new velocity, as well as the subsequent trajectory, depend on the result of the measurement and also on the basis chosen for the measurement. In our work, the time when the ``branching" occurs can be set at any time during the evolution. Identical results would be expected for an experiment where photon $A$ was detected while photon $B$ was still in flight.

\begin{figure}[!htb]
\begin{center}
\includegraphics[width=1\columnwidth]{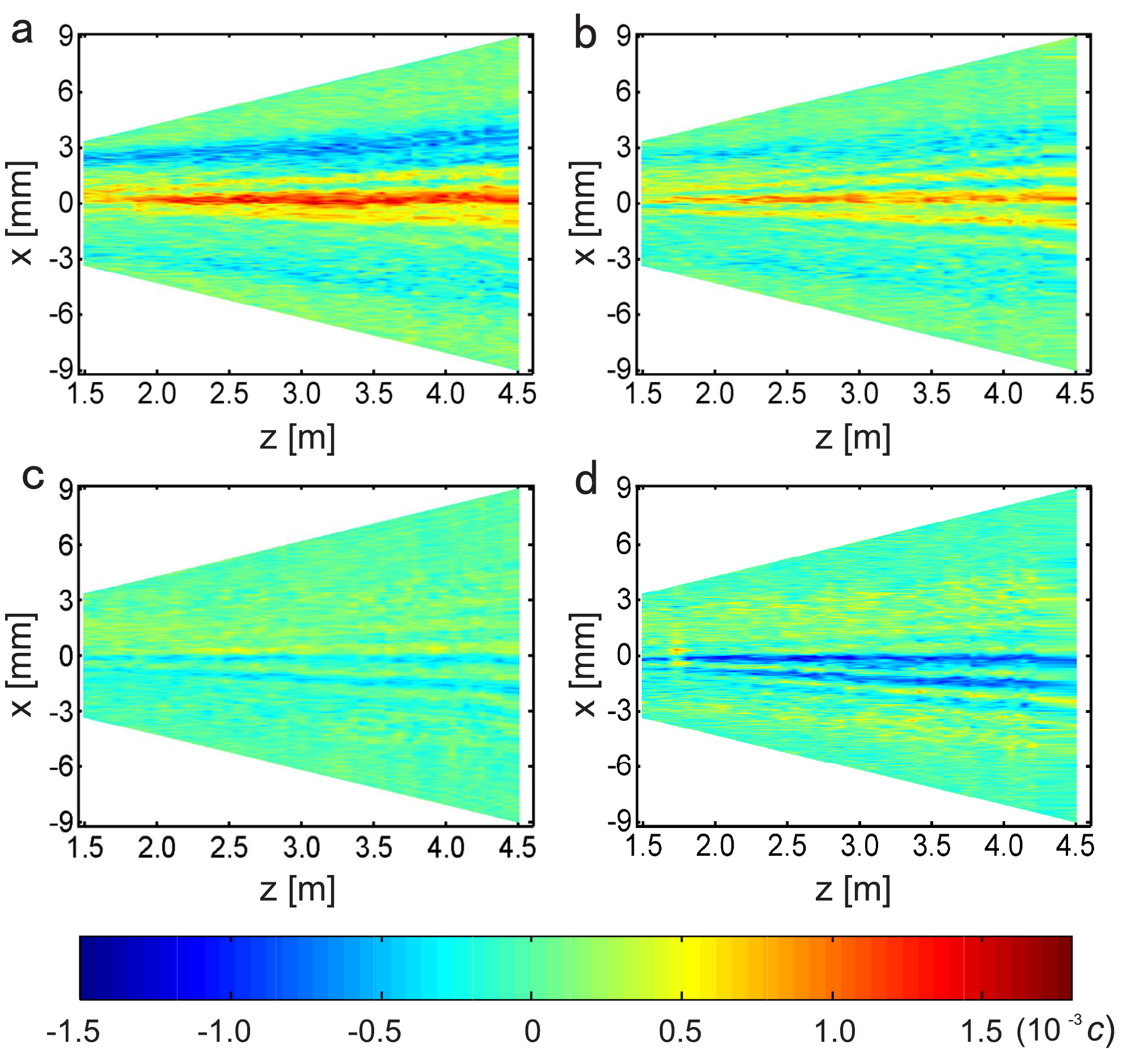}
\caption{The change in velocity of photon B due to projection of photon A. Particle B is in a position $(x,z)$, traveling with transverse velocity $v_x(x,z)$ (without projective-measurement on particle A) and then a remote measurement finds particle A in a state $ \cos\theta \vert H \rangle - \sin\theta \vert V \rangle $. The velocity of particle B is changed and this change is shown for  (a) $\theta=18.5^{\circ} $, (b) $\theta=31.4^{\circ} $, (c) $\theta=45.1^{\circ} $, (d) $\theta=62.9^{\circ} $.}
\label{change}
\end{center}
\end{figure}


In contrast to the common treatment of steering, where the steered system is described by a quantum state, our description is the Bohmian position of a particle. Since this position has a well-defined value, the effect of the steering is self-evident and does not require violation of any inequality. In any case where the Bohmian velocity of a particle is different, after a separate system is measured, one can say the particle was steered. Here, ``steering" is used in the most natural sense of the word: setting a different course. For this to occur the wavefuction (``quantum potential'' or pilot wave in the terminology of BM), have to be steerable, in the sense of \cite{Wiseman2007}, and the particle has to be in a position supporting more than one components of the wavefunction. In our scenario, the double slit scheme, this means where the two wavepackets overlap and interfere with each other. The amount of steering that is performed can be quantified by the change in the velocity, or the momentum transfer \cite{Wiseman}, of particle $B$, due to a measurement on particle $A$. This quantity is shown in Fig. \ref{change} as a plot of the velocity differences between the cases in which remote measurements were carried out on photon $A$ and the cases in the measurement were not. In this sense, our scheme is more intuitive and efficient than the ones which violate steering inequalities to determine the steering. The errors of the velocities, and therefore the corresponding trajectories, which are estimated from the counting statistics, are small and thus are not shown in the figures.

Let us compare now our work to the work done in \cite{ Mahler2016}, where surreal Bohamian trajectories were observed, using a similar experimental setup. From the technical point of view, the main difference was in the type of measurement done on the other photon, i.e. the one whose trajectories were not measured. In \cite{ Mahler2016} the basis of the measurement was $(\vert{H} \rangle- e^{i \phi} \vert{V}\rangle)/ \sqrt{2}$, with varying $\phi$, while we used $\cos \theta \vert{H} \rangle- \sin \theta \vert{V}\rangle$, with varying $\theta $. These can be referred to as phase measurement and amplitude measurement, respectively. This difference resulted in a distinct change in the observed trajectories. In the scenario of phase measurements, the symmetry between the two slits remains and thus no trajectory is crossing the midline $x=0$. In that case, the measuring position of the particle, $x>0$ or $x<0$, can tell us in which slit the particle has passed. For amplitude measurement, the symmetry is broken and the trajectories can start on one side and end up on the other, as seen from Fig. 2.

The analysis of the experiment we present here, and its interpretation, are somewhat opposite to the ones given in \cite{Mahler2016}. The resolution of the surreal trajectories paradox \cite{Englert1992}, is by considering the (non-local) effect of the particle position on the measurement apparatus \cite{Hiley2006}. This apparatus is treated as a vector on the Bloch sphere and its measured evolution is shown to follow Bohmian mechanics.  Our work is aimed at studying the (non-local) effect of the measurement operation on the velocity of the particle. Following the idea of steering, the measured system is not necessarily described by Bohmian mechanics, or quantum mechanics. One can view the choice of measurement basis simply as a decision of the experimentalist of how to align her equipment. The steered system, on the other hand, is analyzed in the full framework of the theory. In contrast to the common use of steering, the relevant theory here is Bohmian mechanics, not quantum mechanics. So while in \cite{Mahler2016}, the peculiarities of BM was highlighted using a feature that is unique to the theory itself, we aimed at studying the theory by employing a more general framework, which can be used to understand other theories, or interpretations, as well.

In addition, we suggest a quantification of the non local effect. It is shown explicitly in Fig. 3, but can also be inferred from Fig. 2 by looking at angle of the trajectory change. It can be difficult to formulate such a quantification in the scenario of \cite{Mahler2016} where a measurement apparatus is evolving continuously along the trajectory of the particle. The picture arising from our work is different: an  measurement instantaneously impacts the velocity of the particle. The change in momentum, mediated instantaneously by the quantum potential, can be a good measure for the non-locality in BM.\\

We should emphasis that since BM is an interpretation of QM, all the results we present agree with the predictions of QM. In particular, the superluminal effect on the Bohmian trajectories cannot be used to transfer information, since the trajectories cannot be obtained for a single particle. The results simply illuminate the implications of adopting such an interpretation, for the underlying physical picture, on the fundamental laws of nature.

\section{Conclusion}
We have experimentally demonstrated non-local steering of Bohmian trajectories. The results show a distinct signature of the non-local nature of BM, by showing the superluminal impact on the velocity of a particle, mediated by the quantum potential, or pilot wave. The phenomenon we observe yields an intuitive picture of the steering: changing the path of a particle from a distance. Nonetheless, it is based on the same mathematical principles as the EPR steering. It could be interesting to see how different hidden variable theories can treat such effects, since any realistic theory that reproduces QM must exhibit strong non-locality.

\appendix
\section*{Appendices}
\section{Methods to measure the coupling strength.}
The relative phase between the horizontal polarization and vertical polarization by weak measurement is expressed as:
\begin{equation}
 \phi(k_{x})=\dfrac{\zeta}{ k }\cdot\langle k_{x}\rangle_{w}=\arcsin(\dfrac{I_{R}-I_{L}}{I_{R}+I_{L}}).
\end{equation}

In our experiment, a piece of 0.7 mm long birefringent calcite, with its optic axis oriented at $42^{\circ}$ with respect to the vertical polarization, is used to perform the weak measurement. The calcite introduces a phase shift $ \phi(k_{x})$ between the ordinary and extraordinary light and a negligibly small walk-off of these two light beams inside the crystal. The birefringent phase shift $ \phi(k_{x})$ depends on the different paths and is approximated as a linear function of $ \langle k_{x}\rangle_{w} $

\begin{equation}
\phi(k_{x})=\zeta\cdot\dfrac{\langle k_{x}\rangle_{w}}{ k }+\phi_{0}.
\end{equation}
If the incident angle  $ \theta_{in}$ of the photon is very small, we can do the following approximation
\begin{equation}
\theta_{in}\approx \sin(\theta_{in})=\dfrac{\langle k_{x}\rangle_{w}}{ k }.
\end{equation}
From the above equations, we can obtain
 \begin{equation}\label{coupling}
\zeta\cdot\theta_{in}+\phi_{0}=\arcsin(\dfrac{I_{R}-I_{L}}{I_{R}+I_{L}}),
\end{equation}
where $\phi_{0}$ is the relative phase with normal incidence. The
interaction strength $ \zeta $ depends on specific parameters of the calcite in the apparatus.

The imaging system with three lens (L1, L2 and L3) is moved. A polarizing beam splitter (PBS) instead of the beam displacer (BD) is used. The intensity of the components with right ($R$) and left ($L$) circular polarization are detected by a power meter from the two ports of PBS. The coupling strength $ \zeta $ in the calcite can be measured and the process is shown as follows:

(1). We use a 808 nm continuous laser light prepared in horizontal polarization to perpendicularly pass through the calcite.

(2). The calcite is then tilted in the normal direction, from  $-0.48^\circ$ to $0.60^\circ $ degrees, in $0.03 ^\circ$ increments in an electrically motorized rotation stage (PR50 CC).

(3). The quarter wave plate (QWP) setting at $45^{\circ}$ and the polarizing beam splitter (PBS) are used to measure the polarization state in the $R/L$ basis, transforming the relative phase between $ \vert H \rangle $ and $ \vert V \rangle $ into an intensity modulation recorded by a power meter.

(4). We use a linear fitting in Origin 8.0 with Eq. (\ref{coupling}) to calculate the local slope $ \zeta $, which is determined to be about 336.

\section{Methods to calculate weak momentum values for different imaging planes}

In our experimental setup, the intensity of the component with horizontal polarization after BD2 corresponds to the projection onto the right-hand circular polarization with the form of $ I_{R}\varpropto[1+\sin(\phi(k_{x}))] $, whereas the intensity of the component with vertical polarization corresponds to the projection onto the left-hand circular polarization with the form of $ I_{L}\varpropto[1-\sin(\phi(k_{x}))] $, which are recorded by the ICCD camera. The pixels on the ICCD where photons are detected correspond to different positions along the direction parallel to the separated direction of the two slits, i.e., the $x$ direction.

The steps to calculate the weak momentum value of transverse momentum $\langle k_{x}\rangle_{w}$ of photon $B$ at the imaging plane when photon $A$ is traced are shown below.

(1). The two intensity distributions of photon $B$ after BD2 are detected on ICCD with photon $A$ projecting to $ \vert H \rangle $. The centers of these two Guassian functions are fitted to be  $ (x_{1} ,x_{3}) $.

(2). The center of these two Gaussian functions are fitted to be $ (x_{2} ,x_{4}) $ when the photon $A$ is projected to $ \vert V \rangle $.

(3). The center position of the right-hand circular polarization is defined as $ R_{c}\equiv(x1+x2)/2 $ and the center position for the left-hand circular polarization is $ L_{x}\equiv(x3+x4)/2 $. The pixels are chosen to be $ [ R_{c}]-160, [ R_{c}]-159.....[ R_{c}]+159, [ R_{c}]+160 $ for the right-hand circular polarization and  $  [L_{c}]-160, [L_{c}]-159.....[L_{c}]+159, [L_{c}]+160 $ for the left-hand circular polarization to reproduce the global propagation behaviour. Here, we defined [    ] as an integral function.

(4). We normalize each density image with the count at each pixel divided by the total counts of all pixels for different polarization. The total counts for right-hand circular polarization and left-hand circular polarization are represented as $S_{R}=\sum^{160}_{-160}N_{R}^{i} $ and $ S_{L}=\sum^{160}_{-160}N_{L}^{i} $ with $ N_{R}^{i} $ and $ N_{L}^{i} $ representing the count at each pixel, respectively. Weak momentum values at each pixel for different imaging planes is calculated as
\begin{center}
$\dfrac{\langle k_{x}\rangle_{w}^{i}}{\vert k \vert}=\dfrac{1}{\zeta}\arcsin[\dfrac{N_{R}^{i}/S_{R}-N_{L}^{i}/S_{L}}{N_{R}^{i}/S_{R}+N_{L}^{i}/S_{L}}]$.
\end{center}
In the reconstructing of trajectories, if a position is not at the center of a pixel, then a Gaussian kernel estimator interpolation between neighbouring momentum values is used.

(5). The corresponding positions (units mm) along the direction parallel to the separation of slits in the equivalent imaging plane without lenses is
\begin{center}
 $ x_{rel}^{z}=\dfrac{(R_{c}-[ R_{c}]+i)*0.013}{\beta}$,
\end{center}
where $\beta$ is the magnification factor and the width of the pixels in the ICCD is 0.013 mm. The cases by projected photon $A$ to $ \cos\theta\vert H \rangle - \sin\theta \vert V \rangle$ can be calculated in a similar way.

\section*{Funding}
This work was supported by the National key research and development program of China (2016YFA0302700); National Natural Science Foundation of China (61327901, 61322506, 11274297, 11325419); the Strategic Priority Research Program (B) of CAS (XDB01030300); the Key Research Program of Frontier Sciences, CAS (QYZDY-SSW-SLH003); the Fundamental Research Funds for the Central Universities (WK2470000020); Youth Innovation Promotion Association and Excellent Young Scientist Program CAS.

\section*{Acknowledgments}
We acknowledge insightful discussions with H. M. Wiseman, M. H. Yung and C. Triola.

\end{document}